\documentclass[12pt]{article}

\usepackage[utf8]{inputenc}
\usepackage[T1]{fontenc}
\usepackage[margin=1in]{geometry}

\usepackage{mathptmx}

\usepackage{setspace}
\usepackage{wrapfig}

\usepackage{titlesec}
\titleformat{\section}[block]{\normalfont\normalsize\bfseries}{\thesection.}{0.5em}{}
\titleformat{\subsection}[block]{\normalfont\normalsize\bfseries}{\thesubsection.}{0.5em}{}
\titleformat{\subsubsection}[block]{\normalfont\normalsize\bfseries}{\thesubsubsection.}{0.5em}{}

\usepackage{graphicx}
\graphicspath{{./}{./figures/}}
\usepackage{float}

\usepackage{amsmath}
\usepackage{amssymb}

\usepackage{booktabs}
\usepackage{threeparttable}
\usepackage{threeparttablex}
\usepackage{makecell}
\usepackage{array}
\usepackage{longtable}
\usepackage{pdflscape}
\usepackage{placeins}
\usepackage{caption}
\usepackage[authoryear,round]{natbib}
\bibpunct{(}{)}{;}{a}{,}{,}

\usepackage{hyperref}
\hypersetup{colorlinks=true,linkcolor=blue,citecolor=blue,urlcolor=blue}

\newcommand{\sym}[1]{\ifmmode^{#1}\else\(^{#1}\)\fi}

\DeclareUnicodeCharacter{2212}{\ensuremath{-}}
\DeclareUnicodeCharacter{0394}{\ensuremath{\Delta}}
\DeclareUnicodeCharacter{2206}{\ensuremath{\Delta}}
\DeclareUnicodeCharacter{00D7}{\ensuremath{\times}}
\DeclareUnicodeCharacter{2248}{\ensuremath{\approx}}
\DeclareUnicodeCharacter{2264}{\ensuremath{\leq}}
\DeclareUnicodeCharacter{2265}{\ensuremath{\geq}}
\DeclareUnicodeCharacter{2009}{\,}
\DeclareUnicodeCharacter{00A0}{~}
\DeclareUnicodeCharacter{2013}{--}
\DeclareUnicodeCharacter{2014}{---}
\DeclareUnicodeCharacter{2019}{'}
\DeclareUnicodeCharacter{201C}{``}
\DeclareUnicodeCharacter{201D}{''}

\makeatletter
\renewcommand\@makefntext[1]{\noindent\@makefnmark\,#1}
\makeatother
\title{Robots and the Public Finance of Disability Insurance\thanks{We are thankful for generous funding from the Social Security
Administration. We thank Marissa Eckrote-Nordland, Mary Hamman, James Murray, and
Michael Stern for helpful comments. We are especially grateful to Pascual Restrepo, who
provided invaluable assistance via numerous email exchanges and the sharing of his code
to facilitate replication of the industrial robot exposure measures from
\citet{acemoglu2020robots}. Any errors are our own. Altindag (altindag@auburn.edu) and
Seals (alan.seals@auburn.edu) are affiliated with Auburn University. Taha
(relcheikhtaha@uwlax.edu) and Nunley (jnunley@uwlax.edu) are affiliated with University of
Wisconsin--La Crosse.}}
\author{Duha T. Altindag, Reem El Cheikh Taha, John M. Nunley, R. Alan Seals}
\date{\ifcase\the\month\or January\or February\or March\or April\or May\or June\or
July\or August\or September\or October\or November\or December\fi\ \the\year}

\begin{document}
\maketitle
\thispagestyle{empty}

\begin{abstract}
\noindent Automation affects public budgets through wages and the tax base, and also through inflows into social insurance. We estimate the effect of industrial robot exposure on Social Security Disability Insurance (SSDI) applications using confidential commuting-zone data and a shift-share design that instruments U.S. exposure with earlier European robot diffusion. One additional robot per 1,000 workers lowers applications by about 8 per 100,000 working-age residents, with the largest declines among workers aged 55 to 64. Employment-to-population ratios do not fall in exposed commuting zones, which weighs against broad local displacement as the sole explanation. A year's flow of averted applications corresponds to about \$3.4 billion in expected SSDI and Medicare obligations, or about \$17,000 per robot-year, in present value rather than realized cash. Displacement adds to social-insurance costs, whereas robot exposure here reduces them, an offset that assessments built only on displacement leave unpriced.
\end{abstract}



\newpage
\thispagestyle{empty}

{\footnotesize\emph{\textbf{Disclaimer: The research reported herein was performed pursuant to a grant from the U.S. Social Security Administration (SSA) funded as part of the Retirement and Disability Research Consortium at the University of Wisconsin--Madison. The opinions and conclusions expressed are solely those of the author(s) and do not represent the opinions or policy of SSA or any agency of the Federal Government. Neither the United States Government nor any agency thereof, nor any of their employees, makes any warranty, express or implied, or assumes any legal liability or responsibility for the accuracy, completeness, or usefulness of the contents of this report. Reference herein to any specific commercial product, process or service by trade name, trademark, manufacturer, or otherwise does not necessarily constitute or imply endorsement, recommendation or favoring by the United States Government or any agency thereof.}}

\textbf{Data Availability.} This paper uses confidential administrative data on SSDI applications from the U.S. Social Security Administration (SSA), proprietary data on industrial robots from the International Federation of Robotics (IFR), and publicly available data from the American Community Survey and Census. The SSA data are confidential and cannot be posted publicly. Researchers can apply for access through SSA. See, for example, \url{https://www.ssa.gov/policy/docs/rsnotes/rsn2009-01.html}. Alan Seals and John Nunley would be available to assist with that process. The IFR data are proprietary and may be obtained directly from the International Federation of Robotics through license or purchase. Nonconfidential replication materials, including code, public-data construction files, and documentation, will be deposited with ICPSR upon publication.

\textbf{Disclosure Statement.} Reem El Cheikh Taha, John M. Nunley, and R. Alan Seals received research support for this project from the U.S. Social Security Administration, Washington, DC, through the Retirement and Disability Research Consortium at the University of Wisconsin--Madison. Duha T. Altindag received no direct project funding for this research. Other than the project support disclosed above, the authors report no consulting fees, retainers, grants, in-kind support, or other financial support relevant to this research. None of the authors serves as an officer, director, or board member of a relevant for-profit or nonprofit entity. The authors report no relevant disclosures involving close relatives or partners. No outside party had the right to review this manuscript prior to circulation. This study received exempt approval from the Auburn University Institutional Review Board (Protocol \#21-533 EX 2111). \par}

\newpage
\setcounter{page}{1}
\section{Introduction}

Each application for Social Security Disability Insurance (SSDI) creates an expected cost for the U.S. federal budget. Application inflows depend partly on labor-market conditions, because workers with health limitations face a choice between continued employment and disability-program entry. Industrial robots can shift that calculus by changing both the physical demands of work and the wages it offers. The sign of this effect is theoretically ambiguous. On the one hand, robots may reduce applications when they substitute for physically demanding tasks or help workers with health limitations remain employed \citep{gihleb2022,filomena2025,kim2026robots}. On the other hand, robots may increase applications when they displace workers \citep{acemoglu2020robots}.

We study the effect of industrial robots on social-insurance inflows using confidential Social Security Administration (SSA) records on SSDI applications. We focus on applications because they capture the program-entry intentions, before SSA adjudicates whether a claim is allowed. Benefit receipt combines the decision to file with the adjudication process, whereas applications measure the full inflow, including claims that are ultimately denied \citep{vonwachter2011,deshpande2019,digiacomo2026}. Applications are fiscally relevant because filings generate administrative costs and, through the probability of award, expected SSDI cash benefits and Medicare obligations \citep{maestas2021}.

We use a shift-share research design that follows \citet{acemoglu2020robots}. Commuting zones with greater preexisting concentration in robot-intensive industries were more exposed to subsequent robot adoption. We instrument U.S. robot exposure using variation in European robot diffusion, interacting industry-level shocks with predetermined local industry shares. Because U.S. robot adoption was concentrated in automotive and related manufacturing during this period, the identifying variation comes from these industries. In robustness analyses, we show that the application decline survives controls for the automotive cycle.

We find that each additional robot per 1,000 workers lowers applications by approximately 8 per 100,000 working-age residents. The estimates are negative across demographic groups and robust to a pre-period placebo test and controls for neighboring-zone exposure. A similar negative pattern appears using an independent, task-based automation measure built from different data and identified by a different instrument, so the finding does not rest on the IFR robot exposure measure alone. Scaling the baseline estimate to observed exposure changes implies roughly 35,000 averted applications per year. Applying the marginal-applicant cost estimates of \citet{maestas2021}, the application response corresponds to about \$3.4 billion per year in expected SSDI and Medicare obligations, or about \$17,000 per robot-year.

The SSDI application response is unlikely to be explained by broad local job loss. Robot-exposed commuting zones show no employment decline over 2004 to 2016, and employment rises for workers aged 55 to 64, the group whose applications fall the most. The employment effect of robots is not the same across the adoption cycle. Early in the sample it reproduces the displacement-era declines of \citet{acemoglu2020robots}, then turns positive in the later diffusion period. This pattern weighs against broad local displacement as the sole explanation for the application response and is consistent with continued labor-market attachment reducing the incentive to file for SSDI.

Existing research on the public-finance consequences of automation emphasizes displacement, wages, and tax-base effects \citep{acemoglu2020tax,guerreiro2022robots,costinot2023,thuemmel2023}. These accounts leave out the rate at which workers apply for disability benefits. \citet{deshpande2026decline} decompose the post-2010 SSDI decline with administrative data and find that it reflects primarily fewer applications and secondarily lower award rates among lower-skilled men whose employment rose, attributable to improved labor-market opportunity. Robot exposure in manufacturing lowers applications without broad local job loss in labor markets where robot adoption is concentrated. We translate the averted applications into expected SSDI and Medicare obligations, a social-insurance term that enters fiscal accounting with the opposite sign from displacement and is left unpriced in standard automation-tax frameworks.

\section{Background and Conceptual Framework}

\subsection{SSDI Applications as a Social Insurance Inflow}

SSDI is a major federal social insurance program whose participation reflects worker health, demographics, labor-market conditions, and program screening standards \citep{autor2003rise,liebman2015}. During the Great Recession, applications rose sharply. \citet{maestas2021} estimate that nearly one million additional SSDI applications were filed during the recession, of which 41.8 percent were awarded benefits. Application rates then declined after 2010 across demographic groups \citep{liu2023}, a decline that individual-level administrative data attribute primarily to fewer applications and secondarily to lower award rates rather than to changes in eligibility \citep{deshpande2026decline}. SSDI inflows shift substantially over short horizons (Appendix Figure~\ref{fig:ssdi_applications}). As of 2022, approximately 9 million individuals received SSDI benefits \citep{ssa2023}. Although the Disability Insurance trust fund is currently projected to remain solvent over the long-run horizon \citep{ssatrustees2023}, marginal applications remain fiscally consequential because they generate administrative costs and, through the probability of award, expected SSDI cash benefits and Medicare obligations.

\subsection{Robots, Work, and Public Budgets}

A large literature studies the labor-market effects of industrial robots. \citet{acemoglu2020robots} use a shift-share design to relate changes in U.S. robot exposure to local employment and wages, and document declines consistent with displacement in robot-exposed commuting zones. In contrast, \citet{dauth2021} find no net employment loss at the local level in Germany, with manufacturing job destruction offset by gains in business services. Firm-level evidence from France and Spain shows adopters expanding at competitors' expense \citep{acemoglu2020competing,koch2021}, so local employment responses depend on general-equilibrium reallocation and product-market competition even when adopting firms expand.

A separate line of research documents the effects of robots on worker health and safety, another channel through which robots can affect SSDI applications. \citet{gihleb2022} show that robot adoption reduces work-related injuries in the United States and Germany, with particularly large effects in manufacturing-intensive regions. Complementing these administrative outcomes, \citet{gunadi2021} find that greater robot penetration is associated with improvements in self-reported health, reductions in work disability, and fewer quits for health reasons, consistent with robots substituting for physically demanding tasks. At the same time, downstream health outcomes are not uniformly positive. \citet{obrien2022} document increases in overdose mortality in U.S. areas with higher robot penetration, while \citet{gihleb2022} find no meaningful effects on suicide. Evidence independent of the European-diffusion instrument used here points the same way, as \citet{filomena2025} show with Italian data that robot adoption reduces fatal and nonfatal workplace accidents, particularly for men in physically demanding jobs. In contrast, \citet{umblijs2025} find that robot adoption in Norwegian manufacturing increases doctor-certified sick leave, with musculoskeletal diagnoses concentrated among blue-collar routine workers and injuries among technical maintenance staff.

Automation also affects public budgets through wages, tax bases, public spending, and the provision of local public goods \citep{spreen2025}, and it can change other social-insurance programs. \citet{kim2026robots} finds that robot exposure in South Korea lowers workers' compensation claims per covered worker, with the largest declines in permanent-disability cases. A workers' compensation claim is largely automatic once a covered worker is injured, so that setting isolates injury incidence, whereas SSDI applications are worker-initiated and medically screened and reflect incidence, insured status, and the decision to file. We study this broader federal inflow and translate the response into expected federal obligations.

\citet{digiacomo2026} study the composition of workers who exit employment following robot exposure and document that a share of displaced workers receive disability benefits. Our outcome differs from theirs. First, we examine SSDI applications rather than benefit receipt. Applications precede adjudication and capture the full inflow into the disability determination system, including claims that are ultimately denied. Second, we estimate unconditional application rates per capita rather than the probability of disability conditional on labor force exit. Third, our estimates capture changes in total inflows to SSDI, not the outcomes of workers who separate from employment. These distinctions imply that aggregate applications can decline even if a subset of displaced workers transitions onto disability. Worker-level evidence on manufacturing shocks documents the opposite-sign case, in which trade-displaced workers move onto disability \citep{autor2014trade} and local disability transfers rise \citep{autor2013china}. The empirical analysis controls for Chinese import competition to separate the robot-exposure estimates from this channel.

\subsection{Channels}

SSDI application inflows depend on workers having, or believing they have, a work-limiting health condition, insured status from recent covered earnings, and the decision to file. Robot adoption can move each. By substituting for strenuous, repetitive, or hazardous tasks, robots may reduce work limitations and lower applications, while a faster work pace or new machine-related hazards could push the other way. By raising or lowering employment and earnings, robots can strengthen or weaken labor-market attachment and insured status. Conditional on health and insured status, the propensity to file responds to economic incentives, and local labor-market shocks move disability participation directly \citep{black2002,autor2003rise}. National evidence attributes much of the recent SSDI application decline to this labor-demand channel \citep{deshpande2026decline}. Examiner-assignment designs show that disability receipt reduces subsequent work \citep{maestas2013,french2014}, while benefit-generosity evidence indicates that disability insurance affects labor supply \citep{gruber2000}.

Because these channels can offset one another, the sign of the aggregate application response is theoretically ambiguous. The timing of applications also need not track the timing of health or employment shocks, since workers may delay filing while they remain employed. Below, we estimate the net reduced-form effect of robot exposure on application inflows over the sample period.\footnote{\citet{messel2019} report a median onset-to-application time of 7.6 months, with longer delays among claimants who continue working after onset.}

\section{Empirical Framework}

\subsection{Estimation Equation}\label{subsec:estimation}

We estimate the effect of commuting-zone exposure to industrial robots on SSDI applications. Our empirical specification follows \citet{acemoglu2020robots}.
\begin{equation}\label{eq:ssdi}
\mathrm{\Delta}A_{\text{gc}} = \alpha + \beta_{g}\mathrm{\Delta}R_{c} + X_{c}^{'}\Psi +
\delta_{d(c)} + \varepsilon_{c}
\end{equation}
The dependent variable \(\mathrm{\Delta}A_{\text{gc}} = A_{\text{gc},2016} - A_{\text{gc},2004}\) is the long difference between 2004 and 2016 in the within-group application rate \(A_{\text{gc}}\), defined as 100 times SSDI applications from subpopulation \(g\) divided by that group's own population in commuting zone \(c\). The subpopulations are defined along sex and age groups (18--34, 35--54, and 55--64), and we estimate equation~\eqref{eq:ssdi} separately for each group \(g\).

\(\Delta R_{c}\) is the change in robot exposure (robots per 1,000 workers) in the commuting zone \(c\) between 2004 and 2016. The vector \(X_{c}\) collects predetermined commuting-zone controls. We build on the control set of \citet{acemoglu2020robots} for comparability with the robot-exposure literature. These controls absorb preexisting differences in local demographic, educational, and industrial composition that could be correlated with both robot adoption and SSDI applications. We also control for exposure to Chinese import competition, cross-commuting-zone variation in administrative stringency, the local unemployment rate, and the employment share of robot-exposed industries, and we include Census division fixed effects (\(\delta_{d(c)}\)).\footnote{The composition controls are the log of population and 2000 shares by sex, race, age (over 65), education (no college, some college, college or professional degree, and master's or doctorate), and manufacturing employment (overall, light manufacturing, and the female share of manufacturing employment). The additional controls are a measure of Chinese import competition (the change in import penetration from 1990 to 2000), the Social Security Administration disability processing time (measured in 2003) capturing administrative stringency, the commuting-zone unemployment rate in 2004, and the employment share of IFR-covered industries.} \(\varepsilon_{c}\) is the error term. The regressions use the 2000 CZ populations as weights, and standard errors are clustered at the state level. From each regression, we obtain the effect of one additional robot per 1,000 workers on the application rate for a group \(g\), indexed by \(\beta_{g}\).

\subsection{Identification}\label{subsec:identification}

OLS estimation of equation~\eqref{eq:ssdi} may be biased if changes in robot exposure are correlated with unobserved CZ-level shocks to health, employment, industry composition, or SSDI filing propensity, or if robot exposure is measured with error. To obtain a causal estimate of \(\beta_{g}\), we follow the shift-share IV strategy of \citet{acemoglu2020robots} and interpret its identifying content through the share-based framework of \citet{goldsmithpinkham2020}.

The instrument uses industry-level European robot diffusion as a source of variation in predicted U.S. robot exposure. The identifying assumption is that, conditional on baseline CZ characteristics and division fixed effects, European industry adoption patterns are unrelated to unobserved U.S. commuting-zone shocks to SSDI application trends. Our instrument, \(\mathrm{\Delta}R_{c}^{\text{EU}}\), is constructed as
\begin{equation}\label{eq:instrument}
\mathrm{\Delta}R_{c}^{\text{EU}} = \sum_{i}
l_{ic,1970}^{\text{US}}\,\mathrm{\Delta}\text{APR}_{i}^{\text{EU}},
\end{equation}
where \(l_{ic,1970}^{\text{US}}\) is the 1970 employment share of industry \(i\) in U.S. commuting zone \(c\), and \(\mathrm{\Delta}\text{APR}_{i}^{\text{EU}}\) is the change in adjusted robot penetration in European industry \(i\) over a pre-U.S. period, 1994 to 2004 (APR is constructed in Section~\ref{subsec:data}). We then estimate equation~\eqref{eq:ssdi} by two-stage least squares (2SLS), with \(\mathrm{\Delta}R_{c}\) instrumented by equation~\eqref{eq:instrument}. Following \citet{acemoglu2020robots}, the instrument uses the earlier 1970 employment shares to capture predetermined exposure, while the endogenous exposure measure in equation~\eqref{eq:robotexposure} uses 1990 shares.

Identification in this design comes from predetermined commuting-zone industry shares interacted with industry-level robot shocks, so the relevant exogeneity condition is on the 1970 shares conditional on baseline characteristics and division fixed effects. Appendix Figures~\ref{fig:appfig_instrument} and~\ref{fig:appfig_shock} show that neither the instrument nor the underlying industry shocks are systematically related to predetermined commuting-zone characteristics.

Between 2004 and 2016, automotive manufacturing absorbed 81 percent of new U.S. industrial robot installations \citep{ifr2019}. This concentration is consistent with the economics of robot adoption, in which high-volume, standardized production makes the fixed costs of integration easier to recover \citep{rosenberg1963,carlsson1984}. The Rotemberg decomposition of our instrument makes this concentration explicit. Automotive carries about 72 percent of the sum of absolute weights and 93 percent of the positive weights, and the effective number of shocks implied by the inverse Herfindahl of the weights is about two rather than the nominal nineteen industries (Appendix Table~\ref{tab:app_rotemberg}).\footnote{An alternative is exposure-robust inference that treats the industry shocks as the level of variation \citep{borusyak2022}. Its large-sample justification relies on many uncorrelated shocks, which the concentration documented here does not provide. For the baseline estimates this approach yields standard errors far smaller than the state-clustered ones, a pattern consistent with understated sampling uncertainty rather than added precision. The same concentration is the condition under which conventional clustered standard errors can also misstate uncertainty in shift-share designs \citep{adao2019}. We report standard errors clustered by state throughout and treat shock concentration as a limitation of inference in this design rather than a problem that clustering by itself resolves.} Thus, following the strategy of \citet{acemoglu2020robots}, our estimates reflect robot exposure within the automotive-centered manufacturing cluster. 

To guard against the possibility that this concentration, rather than robot adoption itself, drives the result, we implement a battery of automotive-specific checks, such as including in the regressions the 2004 to 2016 change in a commuting zone's automotive employment share and allowing a differential trend for each quartile of 1970 automotive penetration. Our baseline estimates survive these and other robustness checks. In addition, we estimate our models with an alternative, task-based measure of automation, the routine employment share \citep{autor2013growth}, identified by an independent shift-share instrument. We obtain directionally similar findings, so our estimates do not rest on the robot measure or the European instrument alone.

\subsection{Data}\label{subsec:data}

Our outcome variable in equation~\eqref{eq:ssdi}, \(\mathrm{\Delta} A_{\text{gc}}\), is the change in the SSDI application rate, measured as applications per 100 working-age individuals, in the commuting zone \(c\) between 2004 and 2016. For each subpopulation \(g\), the denominator is that group's own working-age population. Data on new SSDI applications are confidential and obtained from the Social Security Administration (SSA), while all population counts, including those for demographic subgroups, are drawn from the American Community Survey. Because 2005 is the earliest year in which consistent population measures are available, we use 2005 population counts when constructing application rates for 2004.

We observe the total number of SSDI applications in each commuting zone as well as applications disaggregated by sex, age group, and their interactions. The SSA suppresses commuting-zone-by-year cells with very few applications. After excluding these observations, coverage exceeds 85 percent in all age and age-by-sex subsamples, with missing data concentrated in sparsely populated commuting zones in the Mountain West and West North Central regions. The set of usable commuting zones differs across age groups and specifications. Our baseline analyses focus on a balanced sample of 557 commuting zones for which application rates are observed across all relevant age and sex categories.

The robot exposure variable, \(\mathrm{\Delta} R_{c}\) in equation~\eqref{eq:ssdi}, is constructed using data from the International Federation of Robotics (IFR) on the stock of industrial robots by industry and year, combined with pre-period industry employment shares from the 1990 Census, following \citet{acemoglu2020robots}. For each commuting zone \(c\), we compute
\begin{equation}\label{eq:robotexposure}
\mathrm{\Delta} R_{c} = \sum_{i} l_{ic,1990}^{\text{US}}\,\mathrm{\Delta}\text{APR}_{i}^{\text{US}},
\end{equation}
where \(l_{ic,1990}^{\text{US}}\) denotes the 1990 employment share of industry \(i\) in commuting zone \(c\), and \(\mathrm{\Delta}\text{APR}_{i}^{\text{US}}\) is the change in the U.S. adjusted penetration ratio of robots in industry \(i\) between 2004 and 2016.\footnote{The adjusted penetration ratio for an industry is the change in its stock of robots per 1990 worker, net of the change implied if robot intensity simply tracked industry output growth (measured from EU KLEMS), following \citet{acemoglu2020robots}, equation~15. The exposure measure uses the U.S. adjusted penetration ratio over 2004 to 2016, while the instrument in equation~\eqref{eq:instrument} uses the European adjusted penetration ratio over 1994 to 2004.} The adjusted penetration ratio rescales the change in the stock of industrial robots in an industry by 1990 employment in that industry. The resulting measure is robots per 1,000 workers.

A limitation of the IFR data is that robot stocks are reported jointly for the United States, Mexico, and Canada prior to 2010, with country-specific reporting beginning in 2010. Because the United States accounts for most North American installations, this aggregation likely adds measurement error to pre-2010 exposure. In an auxiliary analysis presented in the appendix, we show that re-estimating our regressions with exposure measures built from the 2010 to 2016 change in U.S.-only stocks (whose correlation with the baseline measure is 0.997) leaves the main results intact.

The commuting-zone covariates in \(X_{c}\), including sex, race, age, educational attainment, industry employment shares, and population size, are drawn from the decennial Censuses and the American Community Surveys. Three of the additional controls come from their original sources. Chinese import competition is from \citet{autor2013china}, and the SSA disability processing time from \citet{kearney2021}. The 2004 commuting-zone unemployment rate is from the Bureau of Labor Statistics' Local Area Unemployment Statistics.

\section{Results}

\subsection{Baseline Estimates}

One additional robot per 1,000 workers lowers total SSDI applications by about 8 per 100,000 working-age residents (Table~\ref{tab:baseline}, column 1 of Panel A), and the first-stage $F$-statistic exceeds 2,000. Appendix Table~\ref{tab:app_firststage} reports the full coefficient set alongside the first-stage, reduced-form, and OLS estimates.

Within group, the decline is largest among older workers (Panel A). For workers aged 55 to 64, one additional robot per 1,000 workers reduces applications by about 13 per 100,000 residents, and the largest single estimate is for women aged 55 to 64, at about 17 per 100,000 (Panel B). These patterns are consistent with an injury- or work-capacity channel of automation, in which robots take over the physically demanding or hazardous tasks that produce work-limiting health conditions \citep{gihleb2022,gunadi2021,filomena2025}. They are also consistent with stronger employment attachment among workers near the disability margin. The response concentrates among older workers, who sit closer to the disability margin and are more likely to translate a reduction in that physical strain into a forgone application.

Robot exposure rose by 2.32 robots per 1,000 workers in the average commuting zone between 2004 and 2016, implying a decline of about 19 applications per 100,000 residents in that zone. Summing the implied response across commuting zones, weighted by their 2016 working-age populations, yields on the order of 35,000 fewer applications per year across the 557 commuting zones in our sample, a response we value in dollar terms below.\footnote{The effect estimate is $-8.2$ per 100{,}000, so the average-zone exposure rise of $2.32$ gives $8.2\times 2.32\approx 19$ per 100{,}000. The sample total is the population-weighted sum $(0.0082/100)\sum_{c}\Delta R_{c}P_{c}\approx 35{,}000$, with $P_{c}$ the 2016 working-age population and $\sum_{c}P_{c}\approx 191$ million.}

\subsection{Incidence of the Aggregate Application Response}

The baseline coefficients measure each subgroup's applications against its own population. They capture within-group propensities to apply rather than the contribution of each group to the aggregate response per resident. Table~\ref{tab:incidence} re-estimates equation~\eqref{eq:ssdi} with applications from each group divided by the total working-age (18--64) population of the commuting zone. Under this normalization the subgroup coefficients sum to the aggregate effect, so each is read as that group's contribution to the change in total SSDI applications per resident. The controls, weights, instrument, and sample are otherwise identical to Table~\ref{tab:baseline}.

Robot exposure lowers total applications per resident by about 8 per 100,000, the same aggregate effect as the baseline (column 1 of Table~\ref{tab:incidence}, identical to column 1 of Table~\ref{tab:baseline}). Prime-age workers aged 35 to 54 account for the largest part of this aggregate, followed by workers aged 55 to 64 and 18 to 34. Men account for most of it. The per-resident rate falls by about 5 per 100,000 for men and 3 for women, so roughly two-thirds of the aggregate response comes from men.

Within sex, the male contribution is spread across all three age groups and is largest for men aged 35 to 54 (Panel B). The female contribution is concentrated at ages 35 to 54 and 55 to 64, while women under 35 contribute little. The aggregate response thus comes primarily from men across the age distribution and from middle-aged and older women. These per-resident contributions are the incidence relevant for the program's finances, since it is the aggregate per-resident inflow, not the within-group application rate, that determines total benefit obligations.

\subsection{Robustness and Scope}\label{subsec:robustness}

The baseline estimates reported in Table~\ref{tab:baseline} are stable across a series of robustness checks. They are similar using the largest available sample for each outcome (Appendix Table~\ref{tab:app_allcz}), a stacked-differences design over the non-overlapping 2004 to 2007 and 2013 to 2016 windows that omits the Great Recession (Appendix Table~\ref{tab:app_stacked}), and a log specification implying that one additional robot per 1,000 workers lowers application rates by about three-quarters of one percent (Appendix Table~\ref{tab:app_dlog}). A placebo that replaces the outcome with the 1994 to 2003 pre-period change shows no differential pre-trend correlated with predicted exposure (Appendix Table~\ref{tab:app_placebo}), and own-zone exposure dominates distance-weighted neighboring-zone exposure (Appendix Table~\ref{tab:app_neighbor}). We also find that the pattern of our estimates is not an artifact of the 2010 change in IFR robot-stock coverage. Building exposure only from the 2010 to 2016 change in U.S.-specific stocks, a measure correlated 0.997 with the baseline, preserves both the older-worker employment increase and the application decline (Appendix Table~\ref{tab:app_epop2010}).

The application decline is not driven by the contemporaneous automotive cycle, even though identification is concentrated in automotive and closely related heavy manufacturing (see the Rotemberg weights in Appendix Table~\ref{tab:app_rotemberg}). Commuting zones with greater 1970 automotive specialization show no differential pre-2004 trend in SSDI applications (Appendix Table~\ref{tab:app_autopre}). Controlling directly for the 2004 to 2016 change in a commuting zone's automotive employment share leaves the estimate near its baseline value, and the automotive employment change itself is statistically indistinguishable from zero (Appendix Table~\ref{tab:app_autoctrl}). Allowing commuting zones in each quartile of 1970 automotive penetration to follow their own trend leaves the estimate essentially unchanged (Appendix Table~\ref{tab:app_autotrend}). 

As an extension, we re-estimate our models with an alternative measure of automation, the routine employment share, the local share of employment in occupations intensive in routine, codifiable tasks. The routine-task content of jobs is a standard proxy for exposure to automation \citep{autorlevymurnane2003,autor2013growth,goos2014}. We measure it as the 2005 commuting-zone share of employment above the 66th percentile of the \citet{deming2017growing} routine-task distribution, instrumented by the \citet{autor2013growth} 1990 leave-one-state-out shift-share. It is correlated about 0.15 with our robot exposure variable across commuting zones. Because the instrument is itself an industry shift-share, we omit the industry-composition controls that are collinear with it. Table~\ref{tab:rsh} shows that this strategy reproduces the negative response across the age and sex subgroups, with the largest declines among older workers. Because the two measures share neither their data nor their instrument, their agreement is independent corroboration of the application response.

\subsection{Employment Attachment and Displacement}

Robot exposure does not lower local employment-to-population ratios over 2004 to 2016, so the application decline does not reflect broad local displacement. Table~\ref{tab:epop} reports 2SLS estimates with the change between 2004 and 2016 in the employment-to-population ratio as the outcome. The aggregate effect is small and positive (column 1), and its confidence interval rules out employment declines larger than about 0.02 percentage points. The estimate rises with age, reaching about 0.28 percentage points at ages 55 to 64, while for the youngest group it is indistinguishable from zero, with a confidence interval that rules out declines larger than about 0.2 percentage points. The gradient holds within sex. The group whose applications fall the most, workers aged 55 to 64, is also the group whose employment is most retained. This co-movement is consistent with continued labor-market attachment rather than exit onto disability. Because applications are measured per resident, a pure displacement or selective-exit account would predict falling employment-to-population ratios in exposed commuting zones. Employment does not fall, which weighs against displacement onto disability or the selective exit of less-healthy workers as the explanation for the application decline.\footnote{In addition, all of our specifications control for local exposure to Chinese import competition \citep{autor2013china,pierce2016}, the shock through which trade-displaced workers move onto disability \citep{autor2014trade}. Therefore, our robot estimates are net of the trade-displacement channel.}

The sign of the employment response shifts over the diffusion path. Appendix Table~\ref{tab:epop_ot} reports the estimate by subperiod, with declines early in the sample, near-zero effects around 2010, and increases thereafter, the years in which the national application decline concentrates. This reconciles the estimates with the negative local employment effects documented for 1990 to 2007 \citep{acemoglu2020robots} and with restructuring that concentrates around recessions \citep{hershbein2018}. The application decline shares this later timing. Windows ending in 2013 show no negative effect, whereas every window ending in 2015 or later yields a negative and statistically significant estimate (Appendix Table~\ref{tab:windows}). Because employment rises rather than falls over this later period, its time profile supports the no-displacement interpretation. These findings suggest that our baseline application results are the effects of the later diffusion period, and earlier episodes with larger displacement could show a different relationship between automation and disability inflows.

\subsection{Budgetary Implications of Averted Applications}

The application response to robot exposure has budgetary implications that can be priced using the marginal-applicant cost parameters of \citet{maestas2021}. They estimate that 41.8 percent of roughly one million recession-induced SSDI applications resulted in awards, with present-value obligations of \$55.7 billion in SSDI cash benefits, \$97.4 billion including Medicare, and \$3.0 billion in claims-processing costs. These figures imply expected SSDI cash and Medicare obligations of about \$97,600 per marginal application. Including processing costs raises the expected fiscal cost to about \$100,600 per application.

Our estimates imply that the observed 2004 to 2016 increase in robot exposure averted roughly 35,000 applications per year across the 557 commuting zones in our sample. At the \citet{maestas2021} benchmark, this application response corresponds to about \$3.4 billion per year in expected SSDI cash and Medicare obligations, or about \$17,000 per robot-year.\footnote{These are expected present-value obligations associated with one year's flow of averted applications. The per-robot-year figure divides the annual total by the roughly 203,000 exposure-implied robots added between 2004 and 2016. Adding claims-processing costs raises the aggregate figure to about \$3.5 billion.} The 557 commuting zones in the estimation sample account for about 72 percent of national SSDI applications in 2004 and 62 percent in 2016, so the dollar figures describe that sample rather than the full country. Sampling error in the baseline coefficient alone implies a 95 percent confidence interval for these figures of about \$1.1 billion to \$5.7 billion in the aggregate, or about \$5,400 to \$27,900 per robot-year.

The main limitation of this calculation is that the award rate and duration profile of robot-averted applications are unobserved. The \citet{maestas2021} benchmark uses a 41.8 percent marginal award rate and an average induced awardee age of 53. Robot-averted applications could differ along both dimensions. We recompute the fiscal value across a grid of alternative award rates and representative awardee ages. For each age, we approximate the present value of an award as a two-percent annuity through conversion to retirement benefits at age 66, calibrated to reproduce the \citet{maestas2021} present value of \$233,500 for an average awardee aged 53.

Table~\ref{tab:fiscal} reports the results. Panel A reports aggregate expected SSDI and Medicare obligations averted per year, and Panel B expresses the same values per robot-year. The full mechanical grid ranges from about \$1.2 billion to \$6.1 billion. The extreme values are unlikely to be realistic because award rates and award durations are not independent. For example, older applicants are typically awarded at higher rates but have shorter expected benefit horizons, while younger applicants have longer horizons but lower award probabilities. A central set of assumptions around the \citet{maestas2021} benchmark, using award rates of 0.35 to 0.45 and representative awardee ages of 50 to 55, implies values between about \$2.4 billion and \$4.4 billion. This award-rate range brackets the cohort award rate for disabled-worker applicants, which fell from about 0.42 in 2005 to 0.31 in 2016 and averaged near 0.36 over the period \citep{ssa2023}.

The averted-obligation flow enters the fiscal accounting of automation with the opposite sign from displacement. Robot-tax analyses typically price the fiscal consequences of lower labor demand, including lost labor-tax revenue and higher transfer payments \citep{acemoglu2020tax,guerreiro2022robots,costinot2023,thuemmel2023}. Our estimates identify an offsetting social-insurance inflow margin, the reduction in expected SSDI and Medicare obligations from robot exposure in automotive-centered manufacturing. While our analysis does not deliver an optimal robot-tax formula, it shows that displacement-focused calibrations leave a relevant fiscal term unpriced.

\section{Conclusion}

Automation affects public budgets through more than wages and the tax base. This paper studies Social Security Disability Insurance (SSDI) application inflows using confidential commuting-zone data and a shift-share design that instruments U.S. robot exposure with earlier European robot diffusion. Each additional robot per 1,000 workers lowers SSDI application inflows by about 8 per 100,000 working-age residents, and employment-to-population ratios do not fall in exposed commuting zones, which weighs against broad local displacement as the only explanation. The identifying variation is concentrated in automotive manufacturing, so the estimates should be interpreted as effects of robot exposure in robot-intensive manufacturing labor markets rather than as effects of all forms of automation. The employment response in these commuting zones is negative early in the sample and non-negative after about 2010, so the estimates speak to the later diffusion period rather than to earlier episodes with larger displacement.

Scaling the baseline estimate to observed exposure changes implies roughly 35,000 averted applications per year. Applying the marginal-applicant cost estimates of \citet{maestas2021}, this response corresponds to about \$3.4 billion per year in expected SSDI and Medicare obligations, or about \$17,000 per robot-year. These figures are expected obligations associated with marginal applications, not realized annual cash savings.

We identify a public-finance externality of automation that is absent from standard fiscal assessments of technological change. The recent national decline in SSDI applications has been traced to improved labor-market opportunity for lower-skilled men, with the source of that opportunity left open \citep{deshpande2026decline}. Our paper points to one possible explanation through robot-intensive labor markets. Specifically, robot exposure lowers applications without broad local job loss, and employment gains are concentrated among older workers, the group whose applications fall most. This pattern is consistent with maintained employment attachment near the disability margin rather than a broad local displacement channel. Studies of automation policy weigh displacement, wages, and the tax base, but leave the offsetting reduction in disability-program obligations we document unpriced \citep{acemoglu2020tax,guerreiro2022robots,costinot2023,thuemmel2023}. Whether similar fiscal effects arise from other automation technologies will depend on the tasks they replace and on how they affect health, work capacity, employment attachment, and filing incentives.

\clearpage
\clearpage
\begin{landscape}
\begin{table}[H]
\caption{Robot Exposure and SSDI Application Inflows}
\label{tab:baseline}
\begin{center}
\setlength{\tabcolsep}{6pt}
\begin{tabular}{l*{6}{c}}
\toprule
\multicolumn{7}{c}{\textbf{Panel A. By age and sex}} \\
& All & 18--34 & 35--54 & 55--64 & Men & Women \\
& (1) & (2) & (3) & (4) & (5) & (6) \\
\midrule
Robot exposure & $-0.0082^{***}$ & $-0.0060^{**}$ & $-0.0061^{**}$ & $-0.0134^{***}$ & $-0.0102^{***}$ & $-0.0063^{***}$ \\
 & $(0.0028)$ & $(0.0028)$ & $(0.0031)$ & $(0.0044)$ & $(0.0033)$ & $(0.0024)$ \\
\addlinespace
Observations & 557 & 557 & 557 & 557 & 557 & 557 \\
\midrule
\multicolumn{7}{c}{\textbf{Panel B. By age within sex}} \\
& \multicolumn{3}{c}{Men} & \multicolumn{3}{c}{Women} \\
\cmidrule(lr){2-4} \cmidrule(lr){5-7}
& 18--34 & 35--54 & 55--64 & 18--34 & 35--54 & 55--64 \\
& (1) & (2) & (3) & (4) & (5) & (6) \\
\midrule
Robot exposure & $-0.0089^{***}$ & $-0.0090^{**}$ & $-0.0094$ & $-0.0031$ & $-0.0037$ & $-0.0167^{***}$ \\
 & $(0.0030)$ & $(0.0036)$ & $(0.0067)$ & $(0.0030)$ & $(0.0028)$ & $(0.0031)$ \\
\addlinespace
Observations & 557 & 557 & 557 & 557 & 557 & 557 \\
\bottomrule
\end{tabular}
\end{center}
{\setstretch{1}\footnotesize\noindent Each column reports a separate two-stage least squares estimate of equation~\eqref{eq:ssdi}. The dependent variable is the change between 2004 and 2016 in SSDI applications per 100 working-age residents in the indicated group. Robot exposure is instrumented by the European shift-share instrument in equation~\eqref{eq:instrument}. All specifications include the \citet{acemoglu2020robots} composition controls, the four additional controls described in Section~\ref{subsec:estimation}, and Census division fixed effects, and are weighted by 2000 commuting-zone population. Standard errors clustered by state in parentheses. The first-stage Kleibergen--Paap $F$ statistic is 2{,}255.9 in every column. $^{*}\,p<0.1$, $^{**}\,p<0.05$, $^{***}\,p<0.01$.\par}
\end{table}
\end{landscape}

\clearpage
\clearpage
\begin{landscape}
\begin{table}[H]
\caption{Incidence of the SSDI Application Response by Age and Sex}
\label{tab:incidence}
\begin{center}
\setlength{\tabcolsep}{6pt}
\begin{tabular}{l*{6}{c}}
\toprule
\multicolumn{7}{c}{\textbf{Panel A. By age and sex}} \\
& All & 18--34 & 35--54 & 55--64 & Men & Women \\
& (1) & (2) & (3) & (4) & (5) & (6) \\
\midrule
Robot exposure & $-0.0082^{***}$ & $-0.0020^{**}$ & $-0.0036^{**}$ & $-0.0025^{**}$ & $-0.0053^{***}$ & $-0.0029^{**}$ \\
 & $(0.0028)$ & $(0.0009)$ & $(0.0015)$ & $(0.0010)$ & $(0.0016)$ & $(0.0013)$ \\
\addlinespace
Observations & 557 & 557 & 557 & 557 & 557 & 557 \\
\midrule
\multicolumn{7}{c}{\textbf{Panel B. By age within sex}} \\
& \multicolumn{3}{c}{Men} & \multicolumn{3}{c}{Women} \\
\cmidrule(lr){2-4} \cmidrule(lr){5-7}
& 18--34 & 35--54 & 55--64 & 18--34 & 35--54 & 55--64 \\
& (1) & (2) & (3) & (4) & (5) & (6) \\
\midrule
Robot exposure & $-0.0017^{***}$ & $-0.0023^{***}$ & $-0.0014^{**}$ & $-0.0004$ & $-0.0014^{**}$ & $-0.0012^{***}$ \\
 & $(0.0004)$ & $(0.0009)$ & $(0.0007)$ & $(0.0005)$ & $(0.0007)$ & $(0.0004)$ \\
\addlinespace
Observations & 557 & 557 & 557 & 557 & 557 & 557 \\
\bottomrule
\end{tabular}
\end{center}
{\setstretch{1}\footnotesize\noindent Each column reports a separate two-stage least squares estimate of equation~\eqref{eq:ssdi} in which applications from the indicated group are divided by the total working-age (18--64) population of the commuting zone. Under this normalization the subgroup coefficients are additive contributions to the change in total SSDI applications per resident, and the column-(1) total equals the aggregate effect in Table~\ref{tab:baseline}. Robot exposure is instrumented by the European shift-share instrument in equation~\eqref{eq:instrument}. All specifications use the same controls, weights, instrument, and balanced 557-commuting-zone sample as Table~\ref{tab:baseline}. Standard errors clustered by state in parentheses. The first-stage Kleibergen--Paap $F$ statistic is 2{,}255.9 in every column. $^{*}\,p<0.1$, $^{**}\,p<0.05$, $^{***}\,p<0.01$.\par}
\end{table}
\end{landscape}

\clearpage
\clearpage
\begin{landscape}
\begin{table}[H]
\caption{An Alternative Automation Measure and Identification}
\label{tab:rsh}
\begin{center}
\setlength{\tabcolsep}{6pt}
\begin{tabular}{l*{6}{c}}
\toprule
\multicolumn{7}{c}{\textbf{Panel A. By age and sex}} \\
& All & 18--34 & 35--54 & 55--64 & Men & Women \\
& (1) & (2) & (3) & (4) & (5) & (6) \\
\midrule
Routine Employment Share & $-0.0371^{***}$ & $-0.0116^{**}$ & $-0.0394^{***}$ & $-0.0715^{***}$ & $-0.0272^{**}$ & $-0.0465^{***}$ \\
& $(0.0097)$ & $(0.0058)$ & $(0.0128)$ & $(0.0190)$ & $(0.0106)$ & $(0.0107)$ \\
\addlinespace
Observations & 557 & 557 & 557 & 557 & 557 & 557 \\
\midrule
\multicolumn{7}{c}{\textbf{Panel B. By age within sex}} \\
& \multicolumn{3}{c}{Men} & \multicolumn{3}{c}{Women} \\
\cmidrule(lr){2-4}\cmidrule(lr){5-7}
& 18--34 & 35--54 & 55--64 & 18--34 & 35--54 & 55--64 \\
& (1) & (2) & (3) & (4) & (5) & (6) \\
\midrule
Routine Employment Share & $-0.0021$ & $-0.0271^{*}$ & $-0.0688^{***}$ & $-0.0213^{***}$ & $-0.0514^{***}$ & $-0.0749^{***}$ \\
& $(0.0083)$ & $(0.0159)$ & $(0.0236)$ & $(0.0077)$ & $(0.0126)$ & $(0.0182)$ \\
\addlinespace
Observations & 557 & 557 & 557 & 557 & 557 & 557 \\
\bottomrule\end{tabular}
\end{center}
{\setstretch{1}\footnotesize\noindent Each column reports a separate two-stage least squares estimate. The dependent variable is the change between 2004 and 2016 in SSDI applications per 100 working-age residents in the indicated group. The treatment is the 2005 routine employment share, the commuting-zone share of employment in occupations above the 66th percentile of the \citet{deming2017growing} routine-task distribution, in percentage points, instrumented by the \citet{autor2013growth} 1990 leave-one-state-out shift-share. Controls, weights, and the balanced 557-commuting-zone sample follow Table~\ref{tab:baseline}, excluding the industry-composition shares. The first-stage Kleibergen--Paap $F$ statistic is 20.1. Standard errors clustered by state are in parentheses. $^{*}\,p<0.1$, $^{**}\,p<0.05$, $^{***}\,p<0.01$.\par}
\end{table}
\end{landscape}

\clearpage
\clearpage
\begin{landscape}
\begin{table}[H]
\caption{Robot Exposure and the Employment-to-Population Ratio}
\label{tab:epop}
\begin{center}
\setlength{\tabcolsep}{6pt}
\begin{tabular}{l*{6}{c}}
\toprule
\multicolumn{7}{c}{\textbf{Panel A. By age and sex}} \\
& All & 18--34 & 35--54 & 55--64 & Men & Women \\
& (1) & (2) & (3) & (4) & (5) & (6) \\
\midrule
Robot exposure & $0.0010^{*}$ & $-0.0004$ & $0.0016^{***}$ & $0.0028^{***}$ & $0.0018^{**}$ & $0.0002$ \\
 & $(0.0006)$ & $(0.0009)$ & $(0.0006)$ & $(0.0008)$ & $(0.0008)$ & $(0.0005)$ \\
\addlinespace
Observations & 557 & 557 & 557 & 557 & 557 & 557 \\
\midrule
\multicolumn{7}{c}{\textbf{Panel B. By age within sex}} \\
& \multicolumn{3}{c}{Men} & \multicolumn{3}{c}{Women} \\
\cmidrule(lr){2-4} \cmidrule(lr){5-7}
& 18--34 & 35--54 & 55--64 & 18--34 & 35--54 & 55--64 \\
& (1) & (2) & (3) & (4) & (5) & (6) \\
\midrule
Robot exposure & $0.0014$ & $0.0018^{**}$ & $0.0029^{**}$ & $-0.0021^{*}$ & $0.0015^{**}$ & $0.0028^{***}$ \\
 & $(0.0014)$ & $(0.0008)$ & $(0.0012)$ & $(0.0011)$ & $(0.0007)$ & $(0.0009)$ \\
\addlinespace
Observations & 557 & 557 & 557 & 557 & 557 & 557 \\
\bottomrule
\end{tabular}
\end{center}
{\setstretch{1}\footnotesize\noindent Each column reports a separate two-stage least squares estimate of equation~\eqref{eq:ssdi} in which the dependent variable is the change between 2004 and 2016 in the employment-to-population ratio of the indicated group. Robot exposure is instrumented by the European shift-share instrument in equation~\eqref{eq:instrument}. All specifications use the same controls, weights, instrument, and balanced 557-commuting-zone sample as Table~\ref{tab:baseline}. Standard errors clustered by state in parentheses. The first-stage Kleibergen--Paap $F$ statistic is 2{,}255.9 in every column. $^{*}\,p<0.1$, $^{**}\,p<0.05$, $^{***}\,p<0.01$.\par}
\end{table}
\end{landscape}

\clearpage
\clearpage
\begin{table}[H]
\caption{The Fiscal Value of Averted SSDI Applications}
\label{tab:fiscal}
\begin{center}
\setlength{\tabcolsep}{8pt}
\begin{tabular}{l*{4}{c}}
\toprule
\multicolumn{5}{c}{\textbf{Panel A. Obligations averted per year (\$ billions)}} \\
& \multicolumn{4}{c}{Awardee age (present value per award)} \\
\cmidrule(lr){2-5}
Award rate & 45 (\$350K) & 50 (\$279K) & 55 (\$201K) & 60 (\$115K) \\
\midrule
0.30 & 3.64 & 2.90 & 2.09 & 1.20 \\
0.35 & 4.24 & 3.39 & 2.44 & 1.40 \\
0.40 & 4.85 & 3.87 & 2.79 & 1.60 \\
0.45 & 5.46 & 4.36 & 3.14 & 1.80 \\
0.50 & 6.06 & 4.84 & 3.49 & 2.00 \\
\midrule
\multicolumn{5}{c}{\textbf{Panel B. Obligations averted per robot-year (\$)}} \\
& \multicolumn{4}{c}{Awardee age} \\
\cmidrule(lr){2-5}
Award rate & 45 & 50 & 55 & 60 \\
\midrule
0.30 & 17{,}900 & 14{,}300 & 10{,}300 & 5{,}900 \\
0.35 & 20{,}900 & 16{,}700 & 12{,}000 & 6{,}900 \\
0.40 & 23{,}900 & 19{,}100 & 13{,}800 & 7{,}900 \\
0.45 & 26{,}900 & 21{,}500 & 15{,}500 & 8{,}900 \\
0.50 & 29{,}900 & 23{,}900 & 17{,}200 & 9{,}800 \\
\bottomrule
\end{tabular}
\end{center}
{\setstretch{1}\footnotesize\noindent This table reports the fiscal value of the SSDI applications averted by robot exposure under alternative assumptions about (i) the fraction of averted applications that would have been awarded benefits, the award rate given in the rows, and (ii) the present value of an award, which depends on the age of a typical awardee and is given in the columns. Panel A reports the aggregate obligations averted per year, and Panel B expresses the same amounts per robot-year. At the \citet{maestas2021} benchmark, a 0.418 award rate at the average induced-awardee age of 53, the aggregate is about \$3.4 billion per year and the per-robot-year figure is about \$16,700. The present value per award, shown in parentheses in the Panel~A column headers, is the discounted stream of cash and Medicare benefits an awardee of the stated age receives until benefits convert to retirement at age 66, at a two percent discount rate, calibrated to the \citet{maestas2021} estimate of \$233{,}500 for their average awardee, who is aged 53.\par}
\end{table}

\clearpage
\bibliographystyle{aea}
\bibliography{references}

\clearpage
\renewcommand{\thetable}{A\arabic{table}}
\renewcommand{\thefigure}{A\arabic{figure}}
\setcounter{table}{0}
\setcounter{figure}{0}
\section*{Appendix}

\begin{figure}[H]
\caption{SSDI Applications Over Time}
\label{fig:ssdi_applications}
\begin{center}
\includegraphics[width=0.8\textwidth]{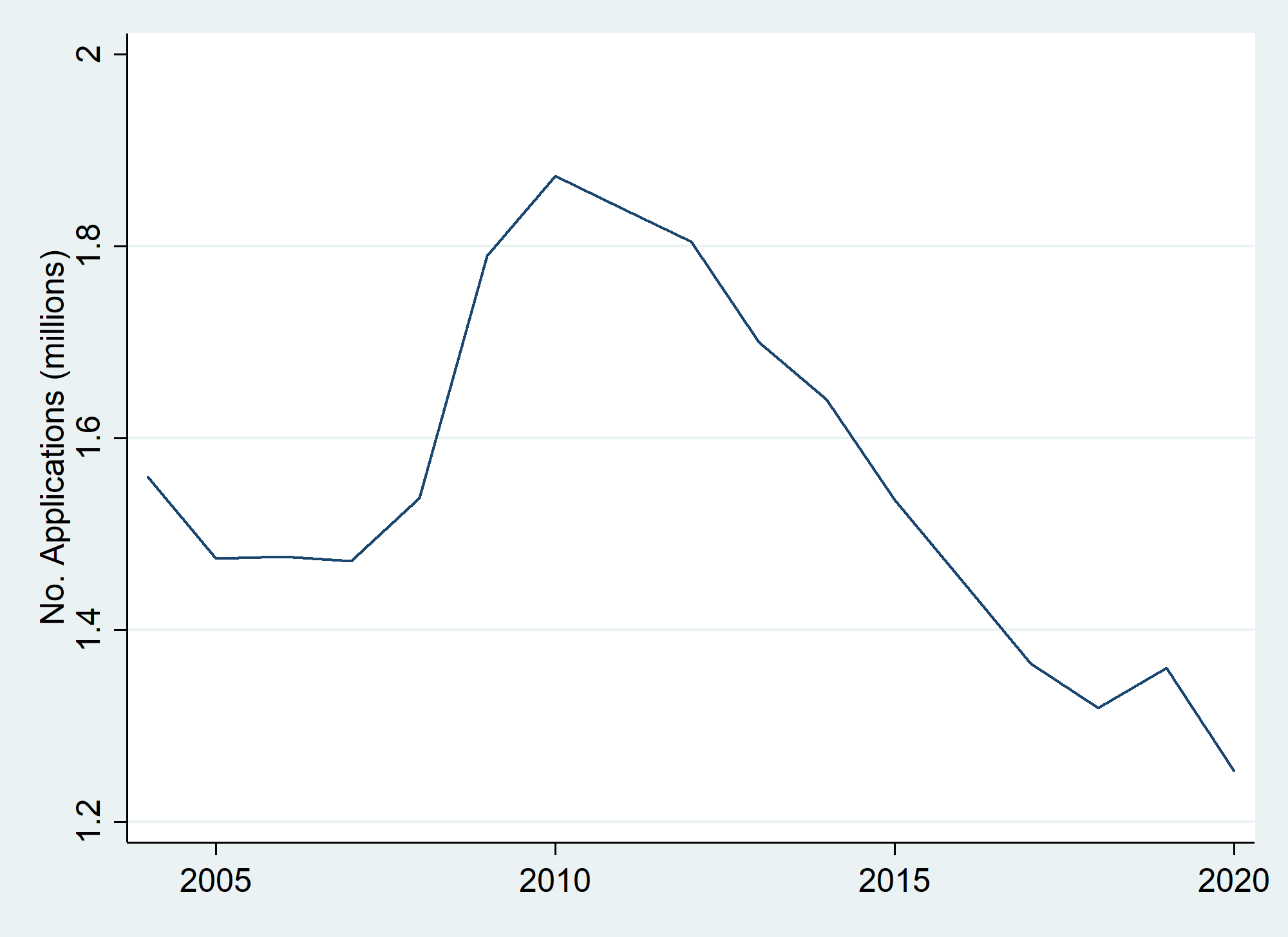}
\end{center}
{\setstretch{1}\footnotesize\noindent The total number of new SSDI applications nationally in each year. The vertical axis is in millions.\par}
\end{figure}
\clearpage
\begin{figure}[H]
\caption{Relationship between the Instrument and Predetermined Commuting-Zone Attributes}
\label{fig:appfig_instrument}
\begin{center}
\includegraphics[width=0.8\textwidth]{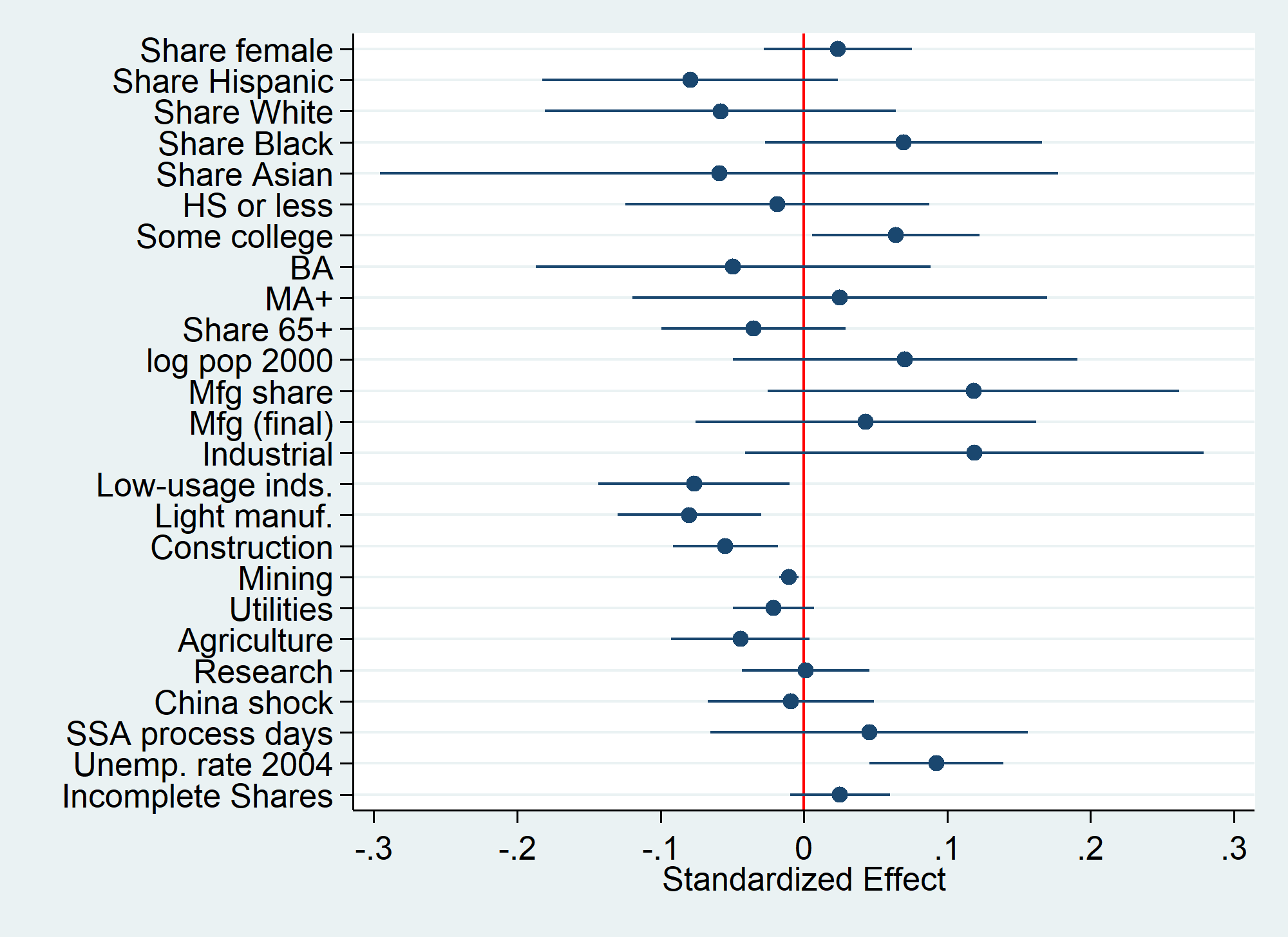}
\end{center}
{\setstretch{1}\footnotesize\noindent Each marker reports the standardized coefficient and its 95 percent confidence interval from a regression of the shift-share robot-exposure instrument on the indicated predetermined commuting-zone attribute, where the attributes are the controls in the vector $X_c$. Coefficients near zero indicate that the instrument is not systematically related to predetermined local characteristics.\par}
\end{figure}
\clearpage
\begin{figure}[H]
\caption{Relationship between the Industry Shocks and Predetermined Commuting-Zone Attributes}
\label{fig:appfig_shock}
\begin{center}
\includegraphics[width=0.8\textwidth]{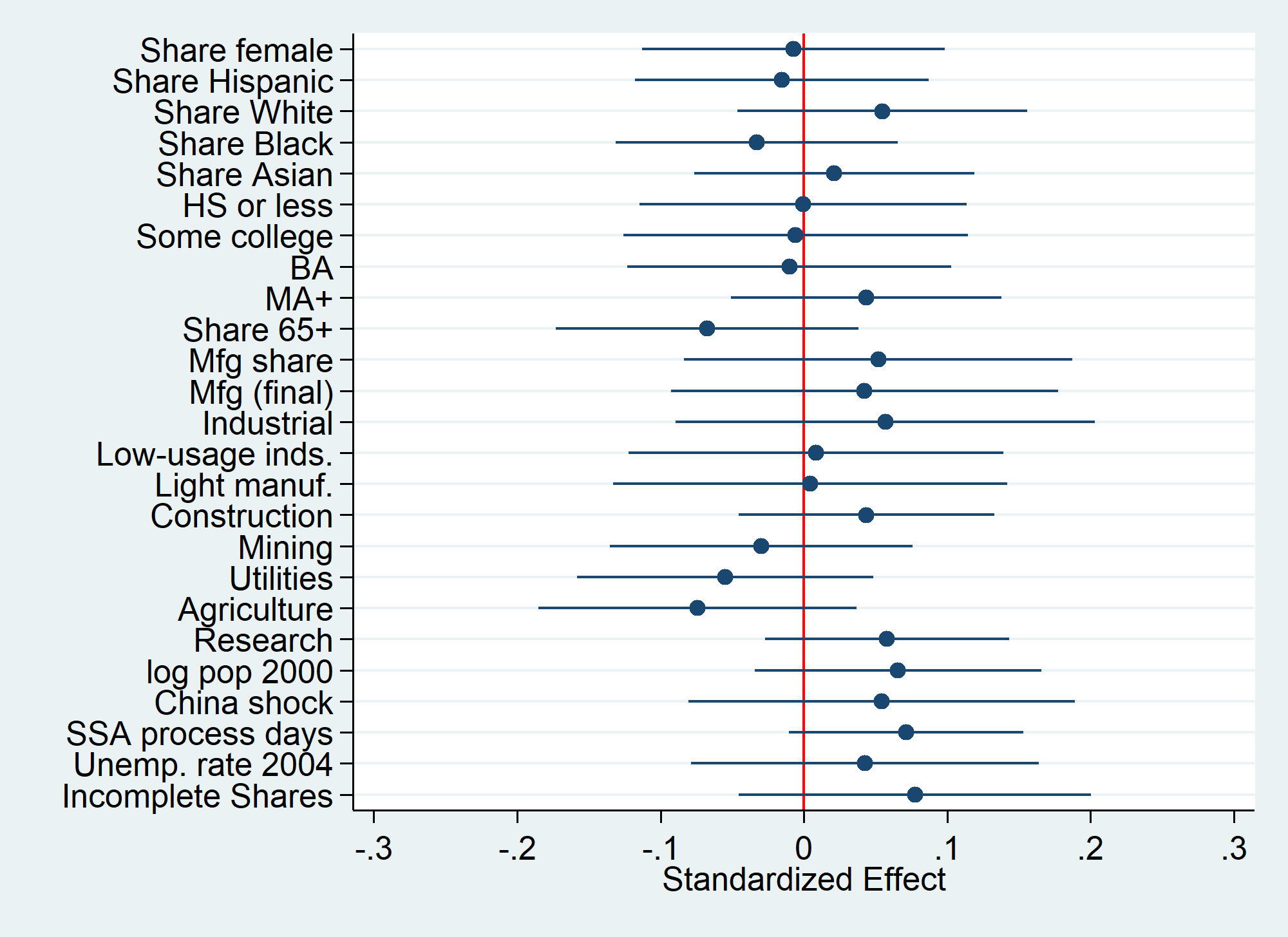}
\end{center}
{\setstretch{1}\footnotesize\noindent Each marker reports the standardized coefficient and its 95 percent confidence interval from the corresponding balance regression for the European industry shocks that underlie the instrument. Coefficients near zero indicate that the shocks are not systematically related to predetermined commuting-zone characteristics.\par}
\end{figure}
\clearpage
\setlength{\tabcolsep}{4pt}
\begin{longtable}{lcccc}
\caption{OLS, First-Stage, Reduced-Form, and 2SLS Estimates}\label{tab:app_firststage}\\
\toprule
& OLS & First Stage & Reduced Form & 2SLS \\
& (1) & (2) & (3) & (4) \\
\midrule
\endfirsthead
\multicolumn{5}{l}{\textit{Table~\thetable\ (continued)}}\\
\toprule
& OLS & First Stage & Reduced Form & 2SLS \\
& (1) & (2) & (3) & (4) \\
\midrule
\endhead
\midrule
\multicolumn{5}{r}{\footnotesize\textit{Continued on next page}}\\
\endfoot
\bottomrule
\endlastfoot
Industrial Robots & $-0.0075^{**}$ &  &  & $-0.0082^{***}$ \\
& $(0.0030)$ &  &  & $(0.0028)$ \\
Female share & $-4.0802^{***}$ & $8.2374$ & $-4.1422^{***}$ & $-4.0750^{***}$ \\
& $(1.0259)$ & $(7.2395)$ & $(1.0285)$ & $(0.9885)$ \\
Hispanic share & $0.2156^{***}$ & $0.0277$ & $0.2153^{***}$ & $0.2155^{***}$ \\
& $(0.0595)$ & $(0.3137)$ & $(0.0597)$ & $(0.0572)$ \\
White share & $-0.1036$ & $-0.1421$ & $-0.1042$ & $-0.1053$ \\
& $(0.1357)$ & $(0.6923)$ & $(0.1375)$ & $(0.1300)$ \\
Black share & $0.2967^{**}$ & $-1.0851$ & $0.3051^{**}$ & $0.2962^{**}$ \\
& $(0.1441)$ & $(0.7765)$ & $(0.1464)$ & $(0.1384)$ \\
Asian share & $-0.5993^{**}$ & $0.9702$ & $-0.6098^{**}$ & $-0.6019^{**}$ \\
& $(0.2689)$ & $(1.7205)$ & $(0.2686)$ & $(0.2573)$ \\
High-school share & $-1.5628^{**}$ & $-4.7650$ & $-1.5487^{**}$ & $-1.5876^{**}$ \\
& $(0.7383)$ & $(3.2216)$ & $(0.7325)$ & $(0.7143)$ \\
Some-college share & $-1.7089^{**}$ & $-1.9166$ & $-1.7084^{**}$ & $-1.7240^{**}$ \\
& $(0.8255)$ & $(3.3580)$ & $(0.8155)$ & $(0.7942)$ \\
College share & $-2.5900^{**}$ & $-10.0470^{**}$ & $-2.5478^{**}$ & $-2.6297^{***}$ \\
& $(0.9880)$ & $(4.8819)$ & $(0.9843)$ & $(0.9549)$ \\
Share aged 65+ & $0.0865$ & $-1.1468$ & $0.0968$ & $0.0875$ \\
& $(0.2731)$ & $(1.6951)$ & $(0.2702)$ & $(0.2643)$ \\
Manufacturing share & $-0.5489$ & $10.0757^{***}$ & $-0.6117$ & $-0.5296$ \\
& $(0.6847)$ & $(2.9523)$ & $(0.7009)$ & $(0.6664)$ \\
Female manuf.\ share & $-2.3139^{***}$ & $14.8832^{**}$ & $-2.4550^{***}$ & $-2.3337^{***}$ \\
& $(0.6917)$ & $(6.7084)$ & $(0.7067)$ & $(0.6686)$ \\
Industrial share & $1.2172^{*}$ & $-3.1405$ & $1.2480^{*}$ & $1.2224^{*}$ \\
& $(0.6869)$ & $(2.2243)$ & $(0.6887)$ & $(0.6589)$ \\
Low-usage share & $-0.3763$ & $-14.0694^{***}$ & $-0.2850$ & $-0.3997$ \\
& $(0.3151)$ & $(2.0233)$ & $(0.2989)$ & $(0.3024)$ \\
Light manuf.\ share & $0.1008$ & $-2.3858$ & $0.1205$ & $0.1011$ \\
& $(0.4027)$ & $(2.7805)$ & $(0.4104)$ & $(0.3892)$ \\
Construction share & $-1.3482^{**}$ & $2.0170$ & $-1.3749^{**}$ & $-1.3585^{**}$ \\
& $(0.6654)$ & $(3.7766)$ & $(0.6547)$ & $(0.6335)$ \\
Utilities share & $-2.7256^{**}$ & $-17.6147^{*}$ & $-2.6197^{**}$ & $-2.7634^{**}$ \\
& $(1.2159)$ & $(9.2750)$ & $(1.1864)$ & $(1.1533)$ \\
Agriculture share & $-0.6185$ & $-2.1275$ & $-0.6080$ & $-0.6253$ \\
& $(0.4428)$ & $(2.4429)$ & $(0.4368)$ & $(0.4246)$ \\
Research share & $-0.9145$ & $-4.6096$ & $-0.8924$ & $-0.9300$ \\
& $(0.5966)$ & $(3.0694)$ & $(0.5905)$ & $(0.5749)$ \\
Log population & $-0.0069$ & $0.0260$ & $-0.0068$ & $-0.0066$ \\
& $(0.0090)$ & $(0.0331)$ & $(0.0090)$ & $(0.0087)$ \\
China shock & $-0.0018$ & $-0.0369$ & $-0.0016$ & $-0.0019$ \\
& $(0.0042)$ & $(0.0344)$ & $(0.0041)$ & $(0.0040)$ \\
SSA processing days & $-0.0000$ & $0.0011$ & $-0.0001$ & $-0.0000$ \\
& $(0.0001)$ & $(0.0008)$ & $(0.0001)$ & $(0.0001)$ \\
Unemployment (2004) & $-0.0210^{***}$ & $0.0453$ & $-0.0212^{***}$ & $-0.0208^{***}$ \\
& $(0.0056)$ & $(0.0390)$ & $(0.0055)$ & $(0.0054)$ \\
Robot stock, 1990 & $-0.0131$ & $-0.0258$ & $-0.0127$ & $-0.0129$ \\
& $(0.0369)$ & $(0.0383)$ & $(0.0367)$ & $(0.0353)$ \\
European robots (IV) &  & $1.3412^{***}$ & $-0.0109^{***}$ &  \\
&  & $(0.0282)$ & $(0.0040)$ &  \\
Constant & $4.1187^{***}$ & $0.1062$ & $4.1354^{***}$ & $4.1363^{***}$ \\
& $(0.8525)$ & $(6.0152)$ & $(0.8490)$ & $(0.8183)$ \\
\addlinespace
Observations & 557 & 557 & 557 & 557 \\
Census division FE & Yes & Yes & Yes & Yes \\
\end{longtable}
\vspace{-1ex}
{\setstretch{1}\footnotesize\noindent Column~1 is OLS, column~2 the first stage (dependent variable is industrial robots), column~3 the reduced form, and column~4 the 2SLS estimate. The dependent variable in columns 1, 3, and 4 is the change in total SSDI applications. State-clustered standard errors in parentheses. $^{*}p<0.1$, $^{**}p<0.05$, $^{***}p<0.01$.\par}
\clearpage
\begin{landscape}
\begin{table}[H]
\caption{Replicating Table 1 Using All Commuting Zones}
\label{tab:app_allcz}
\begin{center}
\setlength{\tabcolsep}{6pt}
\begin{tabular}{l*{6}{c}}
\toprule
\multicolumn{7}{c}{\textit{Panel A}}\\
& All & 18--34 & 35--54 & 55--64 & Men & Women \\
& (1) & (2) & (3) & (4) & (5) & (6) \\
\midrule
Industrial Robots & $-0.0080^{***}$ & $-0.0060^{**}$ & $-0.0057^{*}$ & $-0.0131^{***}$ & $-0.0102^{***}$ & $-0.0063^{***}$ \\
& $(0.0028)$ & $(0.0028)$ & $(0.0031)$ & $(0.0042)$ & $(0.0033)$ & $(0.0024)$ \\
\addlinespace
Observations & 625 & 627 & 692 & 685 & 557 & 557 \\
\midrule
\multicolumn{7}{c}{\textit{Panel B}}\\
& \multicolumn{3}{c}{Men} & \multicolumn{3}{c}{Women} \\
\cmidrule(lr){2-4}\cmidrule(lr){5-7}
& 18--34 & 35--54 & 55--64 & 18--34 & 35--54 & 55--64 \\
& (1) & (2) & (3) & (4) & (5) & (6) \\
\midrule
Industrial Robots & $-0.0089^{***}$ & $-0.0086^{**}$ & $-0.0088$ & $-0.0031$ & $-0.0033$ & $-0.0167^{***}$ \\
& $(0.0030)$ & $(0.0036)$ & $(0.0066)$ & $(0.0030)$ & $(0.0028)$ & $(0.0029)$ \\
\addlinespace
Observations & 559 & 644 & 630 & 559 & 644 & 630 \\
\bottomrule\end{tabular}
\end{center}
{\setstretch{1}\footnotesize\noindent Each column re-estimates the corresponding column of Table~\ref{tab:baseline} on the largest commuting-zone sample available for each outcome rather than the balanced 557-commuting-zone sample. Controls, instrument, weights, and clustering follow Table~\ref{tab:baseline}. $^{*}\,p<0.1$, $^{**}\,p<0.05$, $^{***}\,p<0.01$.\par}
\end{table}
\clearpage
\begin{table}[H]
\caption{Table 1 with Stacked Differences}
\label{tab:app_stacked}
\begin{center}
\setlength{\tabcolsep}{6pt}
\begin{tabular}{l*{6}{c}}
\toprule
\multicolumn{7}{c}{\textit{Panel A}}\\
& All & 18--34 & 35--54 & 55--64 & Men & Women \\
& (1) & (2) & (3) & (4) & (5) & (6) \\
\midrule
Industrial Robots & $-0.0073^{***}$ & $-0.0041^{***}$ & $-0.0048$ & $-0.0145^{***}$ & $-0.0047$ & $-0.0096^{***}$ \\
& $(0.0022)$ & $(0.0014)$ & $(0.0031)$ & $(0.0050)$ & $(0.0030)$ & $(0.0020)$ \\
\addlinespace
Observations & 1119 & 1119 & 1119 & 1119 & 1119 & 1119 \\
\midrule
\multicolumn{7}{c}{\textit{Panel B}}\\
& \multicolumn{3}{c}{Men} & \multicolumn{3}{c}{Women} \\
\cmidrule(lr){2-4}\cmidrule(lr){5-7}
& 18--34 & 35--54 & 55--64 & 18--34 & 35--54 & 55--64 \\
& (1) & (2) & (3) & (4) & (5) & (6) \\
\midrule
Industrial Robots & $-0.0026$ & $-0.0065^{*}$ & $-0.0003$ & $-0.0055^{***}$ & $-0.0033$ & $-0.0269^{***}$ \\
& $(0.0018)$ & $(0.0039)$ & $(0.0095)$ & $(0.0017)$ & $(0.0033)$ & $(0.0034)$ \\
\addlinespace
Observations & 1119 & 1119 & 1119 & 1119 & 1119 & 1119 \\
\bottomrule\end{tabular}
\end{center}
{\setstretch{1}\footnotesize\noindent Each column reports a separate two-stage least squares estimate that stacks the 2004 to 2007 and 2013 to 2016 differences, omitting the Great Recession and early-recovery years. Controls, instrument, and population weights follow Table~\ref{tab:baseline}, and each commuting zone contributes one observation per stacked period difference. Standard errors clustered by state are in parentheses. $^{*}\,p<0.1$, $^{**}\,p<0.05$, $^{***}\,p<0.01$.\par}
\end{table}
\end{landscape}
\begin{table}[H]
\caption{Robot Exposure and SSDI Applications across Outcome Windows}
\label{tab:windows}
\begin{center}
\setlength{\tabcolsep}{8pt}
\renewcommand{\arraystretch}{1.15}
\begin{tabular}{|l|c|c|c|c|}
\hline
 & \multicolumn{4}{c|}{Start year} \\
\cline{2-5}
End year & 2004 & 2005 & 2006 & 2007 \\
\hline
2013 & $0.0050^{*}$ & $0.0027$ & $-0.0026$ & $-0.0026$ \\
 & $(0.0027)$ & $(0.0021)$ & $(0.0024)$ & $(0.0019)$ \\
\hline
2014 & $-0.0023$ & $-0.0045^{*}$ & $-0.0099^{***}$ & $-0.0099^{***}$ \\
 & $(0.0031)$ & $(0.0025)$ & $(0.0026)$ & $(0.0020)$ \\
\hline
2015 & $-0.0058^{**}$ & $-0.0080^{***}$ & $-0.0134^{***}$ & $-0.0134^{***}$ \\
 & $(0.0026)$ & $(0.0019)$ & $(0.0023)$ & $(0.0020)$ \\
\hline
2016 & $-0.0082^{***}$ & $-0.0104^{***}$ & $-0.0157^{***}$ & $-0.0158^{***}$ \\
 & $(0.0028)$ & $(0.0020)$ & $(0.0028)$ & $(0.0024)$ \\
\hline
2017 & $-0.0103^{***}$ & $-0.0126^{***}$ & $-0.0179^{***}$ & $-0.0179^{***}$ \\
 & $(0.0032)$ & $(0.0023)$ & $(0.0030)$ & $(0.0026)$ \\
\hline
2018 & $-0.0081^{***}$ & $-0.0104^{***}$ & $-0.0157^{***}$ & $-0.0157^{***}$ \\
 & $(0.0027)$ & $(0.0021)$ & $(0.0022)$ & $(0.0018)$ \\
\hline
2019 & $-0.0075^{***}$ & $-0.0098^{***}$ & $-0.0151^{***}$ & $-0.0152^{***}$ \\
 & $(0.0027)$ & $(0.0022)$ & $(0.0022)$ & $(0.0018)$ \\
\hline
\end{tabular}
\end{center}
{\setstretch{1}\footnotesize\noindent Each cell is a separate two-stage least squares regression of the change in the total SSDI application rate over the outcome window running from the start year in the column heading to the end year in the row, measured per 100 working-age residents. Robot exposure is held at the 2004 to 2016 measure and instrumented by the European shift-share instrument, with the controls, division fixed effects, population weights, and balanced 557-commuting-zone sample of Table~\ref{tab:baseline}. The baseline specification is the start-2004, end-2016 window. Standard errors clustered by state are in parentheses. $^{*}\,p<0.1$, $^{**}\,p<0.05$, $^{***}\,p<0.01$.\par}
\end{table}
\clearpage
\begin{landscape}
\begin{table}[H]
\caption{Table 1 with Log-Difference Outcomes}
\label{tab:app_dlog}
\begin{center}
\setlength{\tabcolsep}{6pt}
\begin{tabular}{l*{6}{c}}
\toprule
\multicolumn{7}{c}{\textit{Panel A}}\\
& All & 18--34 & 35--54 & 55--64 & Men & Women \\
& (1) & (2) & (3) & (4) & (5) & (6) \\
\midrule
Industrial Robots & $-0.0074^{**}$ & $-0.0093$ & $-0.0024$ & $-0.0092^{***}$ & $-0.0097^{***}$ & $-0.0058^{**}$ \\
& $(0.0032)$ & $(0.0057)$ & $(0.0027)$ & $(0.0028)$ & $(0.0036)$ & $(0.0028)$ \\
\addlinespace
Observations & 557 & 557 & 557 & 557 & 557 & 557 \\
\midrule
\multicolumn{7}{c}{\textit{Panel B}}\\
& \multicolumn{3}{c}{Men} & \multicolumn{3}{c}{Women} \\
\cmidrule(lr){2-4}\cmidrule(lr){5-7}
& 18--34 & 35--54 & 55--64 & 18--34 & 35--54 & 55--64 \\
& (1) & (2) & (3) & (4) & (5) & (6) \\
\midrule
Industrial Robots & $-0.0183^{***}$ & $-0.0052$ & $-0.0062^{*}$ & $-0.0022$ & $-0.0011$ & $-0.0130^{***}$ \\
& $(0.0065)$ & $(0.0033)$ & $(0.0037)$ & $(0.0059)$ & $(0.0024)$ & $(0.0025)$ \\
\addlinespace
Observations & 557 & 557 & 557 & 557 & 557 & 557 \\
\bottomrule\end{tabular}
\end{center}
{\setstretch{1}\footnotesize\noindent Each column reports a separate two-stage least squares estimate of equation~\eqref{eq:ssdi} in which the dependent variable is the log change in the SSDI application rate between 2004 and 2016. Controls, instrument, population weights, and the balanced 557-commuting-zone sample follow Table~\ref{tab:baseline}. Standard errors clustered by state are in parentheses. $^{*}\,p<0.1$, $^{**}\,p<0.05$, $^{***}\,p<0.01$.\par}
\end{table}
\clearpage
\begin{table}[H]
\caption{Table 1 with Pre-Period Outcomes (Change between 1994 and 2003)}
\label{tab:app_placebo}
\begin{center}
\setlength{\tabcolsep}{6pt}
\begin{tabular}{l*{6}{c}}
\toprule
\multicolumn{7}{c}{\textit{Panel A}}\\
& All & 18--34 & 35--54 & 55--64 & Men & Women \\
& (1) & (2) & (3) & (4) & (5) & (6) \\
\midrule
Industrial Robots & $0.0025$ & $0.0073$ & $-0.0015$ & $-0.0029$ & $0.0004$ & $0.0044$ \\
& $(0.0056)$ & $(0.0060)$ & $(0.0060)$ & $(0.0049)$ & $(0.0080)$ & $(0.0035)$ \\
\addlinespace
Observations & 541 & 541 & 541 & 541 & 541 & 541 \\
\midrule
\multicolumn{7}{c}{\textit{Panel B}}\\
& \multicolumn{3}{c}{Men} & \multicolumn{3}{c}{Women} \\
\cmidrule(lr){2-4}\cmidrule(lr){5-7}
& 18--34 & 35--54 & 55--64 & 18--34 & 35--54 & 55--64 \\
& (1) & (2) & (3) & (4) & (5) & (6) \\
\midrule
Industrial Robots & $0.0075$ & $-0.0043$ & $-0.0077$ & $0.0067$ & $0.0012$ & $0.0017$ \\
& $(0.0076)$ & $(0.0088)$ & $(0.0083)$ & $(0.0046)$ & $(0.0038)$ & $(0.0029)$ \\
\addlinespace
Observations & 541 & 541 & 541 & 541 & 541 & 541 \\
\bottomrule\end{tabular}
\end{center}
{\setstretch{1}\footnotesize\noindent Each column reports a separate two-stage least squares estimate of equation~\eqref{eq:ssdi} in which the dependent variable is replaced by the pre-period change in the SSDI application rate between 1994 and 2003. Controls, instrument, and population weights follow Table~\ref{tab:baseline}. The sample is the balanced Table~\ref{tab:baseline} commuting zones for which SSDI application rates are observed in both 1994 and 2003, which reduces the count to 541. Standard errors clustered by state are in parentheses. $^{*}\,p<0.1$, $^{**}\,p<0.05$, $^{***}\,p<0.01$.\par}
\end{table}
\clearpage
\begin{table}[H]
\caption{Table 1 Controlling for Neighboring-CZ Robots}
\label{tab:app_neighbor}
\begin{center}
\setlength{\tabcolsep}{6pt}
\begin{tabular}{l*{6}{c}}
\toprule
\multicolumn{7}{c}{\textit{Panel A}}\\
& All & 18--34 & 35--54 & 55--64 & Men & Women \\
& (1) & (2) & (3) & (4) & (5) & (6) \\
\midrule
Industrial Robots & $-0.0067^{***}$ & $-0.0041$ & $-0.0047$ & $-0.0157^{***}$ & $-0.0087^{***}$ & $-0.0047^{**}$ \\
& $(0.0026)$ & $(0.0026)$ & $(0.0032)$ & $(0.0043)$ & $(0.0029)$ & $(0.0024)$ \\
Neighbor Robots & $-0.0502$ & $-0.0638$ & $-0.0458$ & $0.0785$ & $-0.0499$ & $-0.0525$ \\
& $(0.0520)$ & $(0.0418)$ & $(0.0690)$ & $(0.0693)$ & $(0.0637)$ & $(0.0450)$ \\
\addlinespace
Observations & 557 & 557 & 557 & 557 & 557 & 557 \\
\midrule
\multicolumn{7}{c}{\textit{Panel B}}\\
& \multicolumn{3}{c}{Men} & \multicolumn{3}{c}{Women} \\
\cmidrule(lr){2-4}\cmidrule(lr){5-7}
& 18--34 & 35--54 & 55--64 & 18--34 & 35--54 & 55--64 \\
& (1) & (2) & (3) & (4) & (5) & (6) \\
\midrule
Industrial Robots & $-0.0064^{**}$ & $-0.0082^{**}$ & $-0.0130^{**}$ & $-0.0018$ & $-0.0017$ & $-0.0178^{***}$ \\
& $(0.0026)$ & $(0.0036)$ & $(0.0061)$ & $(0.0031)$ & $(0.0031)$ & $(0.0038)$ \\
Neighbor Robots & $-0.0838^{*}$ & $-0.0268$ & $0.1216$ & $-0.0440$ & $-0.0652$ & $0.0388$ \\
& $(0.0462)$ & $(0.0849)$ & $(0.0864)$ & $(0.0435)$ & $(0.0593)$ & $(0.0680)$ \\
\addlinespace
Observations & 557 & 557 & 557 & 557 & 557 & 557 \\
\bottomrule\end{tabular}
\end{center}
{\setstretch{1}\footnotesize\noindent Each column reports a separate two-stage least squares estimate that augments equation~\eqref{eq:ssdi} with distance-weighted robot exposure in other commuting zones. The remaining controls, instrument, weights, and the balanced 557-commuting-zone sample follow Table~\ref{tab:baseline}. Standard errors clustered by state are in parentheses. $^{*}\,p<0.1$, $^{**}\,p<0.05$, $^{***}\,p<0.01$.\par}
\end{table}
\clearpage
\end{landscape}
\begin{table}[H]
\caption{Robustness to the 2010 Change in IFR Robot-Stock Coverage}
\label{tab:app_epop2010}
\begin{center}
\setlength{\tabcolsep}{10pt}
\begin{tabular}{lccc}
\toprule
& All & Ages 35--54 & Ages 55--64 \\
& (1) & (2) & (3) \\
\midrule
\multicolumn{4}{l}{\textit{Panel A. Employment-to-population ratio, 2004--2016 change}} \\
Robot exposure, 2004--2016 (baseline) & $0.0010^{*}$ & $0.0016^{***}$ & $0.0028^{***}$ \\
 & $(0.0006)$ & $(0.0006)$ & $(0.0008)$ \\
Robot exposure, 2010--2016 (U.S.-specific) & $0.0019^{*}$ & $0.0033^{***}$ & $0.0057^{***}$ \\
 & $(0.0012)$ & $(0.0012)$ & $(0.0017)$ \\
\midrule
\multicolumn{4}{l}{\textit{Panel B. Employment-to-population ratio, 2010--2016 change}} \\
Robot exposure, 2010--2016 (U.S.-specific) & $0.0069^{***}$ & $0.0026$ & $0.0096^{***}$ \\
 & $(0.0009)$ & $(0.0016)$ & $(0.0012)$ \\
\midrule
\multicolumn{4}{l}{\textit{Panel C. SSDI applications, 2004--2016 change}} \\
Robot exposure, 2004--2016 (baseline) & $-0.0082^{***}$ & $-0.0061^{**}$ & $-0.0134^{***}$ \\
 & $(0.0028)$ & $(0.0031)$ & $(0.0044)$ \\
Robot exposure, 2010--2016 (U.S.-specific) & $-0.0164^{***}$ & $-0.0123^{**}$ & $-0.0270^{***}$ \\
 & $(0.0056)$ & $(0.0062)$ & $(0.0088)$ \\
\bottomrule
\end{tabular}
\end{center}
{\setstretch{1}\footnotesize\noindent The International Federation of Robotics reports U.S.-specific robot stocks only from 2010, and earlier stocks are a combined United States, Canada, and Mexico total. The baseline exposure measure spans 2004 to 2016 and crosses this reporting change, while the alternative measure uses only the 2010 to 2016 change in U.S.-specific stocks. The two measures are correlated 0.997 across commuting zones. Each cell is a separate two-stage least squares estimate instrumented by the European shift-share instrument, with the controls, division fixed effects, population weights, and balanced 557-commuting-zone sample of Table~\ref{tab:baseline}. Panel B regresses the 2010 to 2016 change in the employment-to-population ratio on the 2010 to 2016 exposure measure, so both sides fall entirely after the reporting change. The first-stage Kleibergen--Paap $F$ statistic is 2{,}255.9 with the 2004 to 2016 exposure and 1{,}740.1 with the 2010 to 2016 exposure. Standard errors clustered by state are in parentheses. $^{*}\,p<0.1$, $^{**}\,p<0.05$, $^{***}\,p<0.01$.\par}
\end{table}
\clearpage
\begin{table}[H]
\caption{Rotemberg Weights by Industry}
\label{tab:app_rotemberg}
\begin{center}
\setlength{\tabcolsep}{10pt}
\begin{tabular}{lc}
\toprule
Industry & Weight $\hat{\alpha}$ \\
\midrule
Automotive & 1.295 \\
Basic Metals & 0.085 \\
Minerals & 0.012 \\
Services & 0.000 \\
Agriculture & -0.001 \\
Utilities & -0.001 \\
Construction & -0.002 \\
Education and Research & -0.004 \\
Paper and Printing & -0.008 \\
Mining & -0.012 \\
Shipbuilding and Aerospace & -0.012 \\
Metal products & -0.014 \\
Textiles & -0.015 \\
Wood and Furniture & -0.028 \\
Plastics and Chemicals & -0.035 \\
Food and Beverages & -0.050 \\
Miscellaneous & -0.050 \\
Industrial Machinery & -0.083 \\
Electronics & -0.086 \\
\bottomrule\end{tabular}
\end{center}
{\setstretch{1}\footnotesize\noindent Rotemberg weights for the IFR-19 industries following \citet{goldsmithpinkham2020}.\par}
\end{table}
\clearpage
\begin{landscape}
\begin{longtable}{l*{6}{c}}
\caption{1970 Automotive Share and Pre-Period SSDI Trends}\label{tab:app_autopre}\\
\toprule
\endfirsthead
\multicolumn{7}{l}{\textit{Table~\thetable\ (continued)}}\\
\toprule
\endhead
\midrule
\multicolumn{7}{r}{\footnotesize\textit{Continued on next page}}\\
\endfoot
\bottomrule
\endlastfoot
\multicolumn{7}{c}{\textit{Panel A. Division FEs only}}\\
& All & 18--34 & 35--54 & 55--64 & Men & Women \\
& (1) & (2) & (3) & (4) & (5) & (6) \\
\midrule
1970 Auto Share & $0.0241$ & $0.0400$ & $0.0059$ & $0.0077$ & $0.0173$ & $0.0297$ \\
& $(0.0520)$ & $(0.0460)$ & $(0.0599)$ & $(0.0438)$ & $(0.0655)$ & $(0.0391)$ \\
\addlinespace
Observations & 541 & 541 & 541 & 541 & 541 & 541 \\
\midrule
\multicolumn{7}{c}{\textit{Panel B. Division FEs only}}\\
& \multicolumn{3}{c}{Men} & \multicolumn{3}{c}{Women} \\
\cmidrule(lr){2-4}\cmidrule(lr){5-7}
& 18--34 & 35--54 & 55--64 & 18--34 & 35--54 & 55--64 \\
& (1) & (2) & (3) & (4) & (5) & (6) \\
\midrule
1970 Auto Share & $0.0296$ & $0.0005$ & $0.0022$ & $0.0479$ & $0.0105$ & $0.0147$ \\
& $(0.0543)$ & $(0.0766)$ & $(0.0616)$ & $(0.0379)$ & $(0.0443)$ & $(0.0281)$ \\
\addlinespace
Observations & 541 & 541 & 541 & 541 & 541 & 541 \\
\midrule
\multicolumn{7}{c}{\textit{Panel C. Complete controls}}\\
& All & 18--34 & 35--54 & 55--64 & Men & Women \\
& (1) & (2) & (3) & (4) & (5) & (6) \\
\midrule
1970 Auto Share & $0.0124$ & $0.0341$ & $-0.0060$ & $-0.0102$ & $0.0024$ & $0.0215$ \\
& $(0.0279)$ & $(0.0303)$ & $(0.0303)$ & $(0.0226)$ & $(0.0402)$ & $(0.0173)$ \\
\addlinespace
Observations & 541 & 541 & 541 & 541 & 541 & 541 \\
\midrule
\multicolumn{7}{c}{\textit{Panel D. Complete controls}}\\
& \multicolumn{3}{c}{Men} & \multicolumn{3}{c}{Women} \\
\cmidrule(lr){2-4}\cmidrule(lr){5-7}
& 18--34 & 35--54 & 55--64 & 18--34 & 35--54 & 55--64 \\
& (1) & (2) & (3) & (4) & (5) & (6) \\
\midrule
1970 Auto Share & $0.0343$ & $-0.0183$ & $-0.0330$ & $0.0320$ & $0.0058$ & $0.0124$ \\
& $(0.0389)$ & $(0.0447)$ & $(0.0385)$ & $(0.0230)$ & $(0.0183)$ & $(0.0139)$ \\
\addlinespace
Observations & 541 & 541 & 541 & 541 & 541 & 541 \\
\end{longtable}
\vspace{-1ex}
{\setstretch{1}\footnotesize\noindent Each column reports a regression of the pre-period change in the SSDI application rate between 1994 and 2003 on the commuting zone's 1970 automotive employment share. Panels A and B include division fixed effects only, and panels C and D add the full control set of Table~\ref{tab:baseline}. The sample is the balanced Table~\ref{tab:baseline} commuting zones for which SSDI application rates are observed in both 1994 and 2003, which reduces the count to 541. Standard errors clustered by state are in parentheses. $^{*}\,p<0.1$, $^{**}\,p<0.05$, $^{***}\,p<0.01$.\par}
\clearpage
\begin{table}[H]
\caption{Table 1 Controlling for the Growth in Automotive Sector Employment}
\label{tab:app_autoctrl}
\begin{center}
\setlength{\tabcolsep}{6pt}
\begin{tabular}{l*{6}{c}}
\toprule
\multicolumn{7}{c}{\textit{Panel A}}\\
& All & 18--34 & 35--54 & 55--64 & Men & Women \\
& (1) & (2) & (3) & (4) & (5) & (6) \\
\midrule
Industrial Robots & $-0.0087^{***}$ & $-0.0068^{**}$ & $-0.0059^{*}$ & $-0.0151^{***}$ & $-0.0105^{***}$ & $-0.0070^{***}$ \\
& $(0.0027)$ & $(0.0028)$ & $(0.0033)$ & $(0.0042)$ & $(0.0032)$ & $(0.0024)$ \\
Auto Share & $-0.5111$ & $-0.7746$ & $0.1841$ & $-1.6052$ & $-0.2662$ & $-0.7298$ \\
& $(1.1471)$ & $(0.7171)$ & $(1.6383)$ & $(1.5216)$ & $(1.2597)$ & $(1.1123)$ \\
\addlinespace
Observations & 557 & 557 & 557 & 557 & 557 & 557 \\
\midrule
\multicolumn{7}{c}{\textit{Panel B}}\\
& \multicolumn{3}{c}{Men} & \multicolumn{3}{c}{Women} \\
\cmidrule(lr){2-4}\cmidrule(lr){5-7}
& 18--34 & 35--54 & 55--64 & 18--34 & 35--54 & 55--64 \\
& (1) & (2) & (3) & (4) & (5) & (6) \\
\midrule
Industrial Robots & $-0.0094^{***}$ & $-0.0086^{**}$ & $-0.0108^{*}$ & $-0.0042$ & $-0.0036$ & $-0.0186^{***}$ \\
& $(0.0031)$ & $(0.0037)$ & $(0.0066)$ & $(0.0029)$ & $(0.0031)$ & $(0.0032)$ \\
Auto Share & $-0.4548$ & $0.3728$ & $-1.4180$ & $-1.0597$ & $0.0220$ & $-1.8650$ \\
& $(0.9387)$ & $(1.7443)$ & $(1.9414)$ & $(0.6632)$ & $(1.6231)$ & $(1.3643)$ \\
\addlinespace
Observations & 557 & 557 & 557 & 557 & 557 & 557 \\
\bottomrule\end{tabular}
\end{center}
{\setstretch{1}\footnotesize\noindent Each column reports a separate two-stage least squares estimate that augments equation~\eqref{eq:ssdi} with the long change in the commuting zone's automotive employment share. The remaining controls, instrument, weights, and the balanced 557-commuting-zone sample follow Table~\ref{tab:baseline}. Standard errors clustered by state are in parentheses. $^{*}\,p<0.1$, $^{**}\,p<0.05$, $^{***}\,p<0.01$.\par}
\end{table}
\clearpage
\begin{table}[H]
\caption{Robot Exposure and SSDI Applications, Controlling for Differential Trends by 1970 Automotive Penetration}
\label{tab:app_autotrend}
\begin{center}
\setlength{\tabcolsep}{6pt}
\begin{tabular}{l*{6}{c}}
\toprule
\multicolumn{7}{c}{\textbf{Panel A. By age and sex}} \\
& All & 18--34 & 35--54 & 55--64 & Men & Women \\
& (1) & (2) & (3) & (4) & (5) & (6) \\
\midrule
Robot exposure & $-0.0088^{***}$ & $-0.0062^{**}$ & $-0.0070^{**}$ & $-0.0153^{***}$ & $-0.0111^{***}$ & $-0.0066^{***}$ \\
 & $(0.0024)$ & $(0.0024)$ & $(0.0029)$ & $(0.0038)$ & $(0.0028)$ & $(0.0022)$ \\
\addlinespace
Observations & 557 & 557 & 557 & 557 & 557 & 557 \\
\midrule
\multicolumn{7}{c}{\textbf{Panel B. By age within sex}} \\
& \multicolumn{3}{c}{Men} & \multicolumn{3}{c}{Women} \\
\cmidrule(lr){2-4} \cmidrule(lr){5-7}
& 18--34 & 35--54 & 55--64 & 18--34 & 35--54 & 55--64 \\
& (1) & (2) & (3) & (4) & (5) & (6) \\
\midrule
Robot exposure & $-0.0088^{***}$ & $-0.0106^{***}$ & $-0.0129^{**}$ & $-0.0037$ & $-0.0039$ & $-0.0171^{***}$ \\
 & $(0.0027)$ & $(0.0032)$ & $(0.0055)$ & $(0.0025)$ & $(0.0028)$ & $(0.0031)$ \\
\addlinespace
Observations & 557 & 557 & 557 & 557 & 557 & 557 \\
\bottomrule
\end{tabular}
\end{center}
{\setstretch{1}\footnotesize\noindent Each column reports a separate two-stage least squares estimate of equation~\eqref{eq:ssdi}, augmenting the baseline specification with indicators for quartiles of the commuting zone's 1970 automotive employment share so that commuting zones in each quartile follow their own trend. The dependent variable is the change between 2004 and 2016 in SSDI applications per 100 working-age residents in the indicated group. All other controls, division fixed effects, population weights, and the balanced 557-commuting-zone sample follow Table~\ref{tab:baseline}. Standard errors clustered by state are in parentheses. The first-stage Kleibergen--Paap $F$ statistic is 3{,}039.2 in every column. $^{*}\,p<0.1$, $^{**}\,p<0.05$, $^{***}\,p<0.01$.\par}
\end{table}
\end{landscape}
\begin{table}[H]
\caption{Robot Exposure and the Employment-to-Population Ratio across Subperiods}
\label{tab:epop_ot}
\begin{center}
\setlength{\tabcolsep}{10pt}
\begin{tabular}{lcccc}
\toprule
& 2004--2007 & 2007--2010 & 2010--2013 & 2013--2016 \\
& (1) & (2) & (3) & (4) \\
\midrule
All & $-0.0019^{***}$ & $-0.0005$ & $0.0014^{***}$ & $0.0020^{***}$ \\
 & $(0.0004)$ & $(0.0005)$ & $(0.0003)$ & $(0.0004)$ \\
Ages 55--64 & $-0.0026^{***}$ & $-0.0010$ & $0.0016^{***}$ & $0.0032^{***}$ \\
 & $(0.0008)$ & $(0.0010)$ & $(0.0005)$ & $(0.0007)$ \\
\bottomrule
\end{tabular}
\end{center}
{\setstretch{1}\footnotesize\noindent Each cell is a separate two-stage least squares estimate of the effect of robot exposure on the change in the employment-to-population ratio over the indicated subperiod. Robot exposure is held at the 2004 to 2016 measure and instrumented by the European shift-share instrument, with the controls, division fixed effects, population weights, and balanced 557-commuting-zone sample of Table~\ref{tab:baseline}. Standard errors clustered by state are in parentheses. The first-stage Kleibergen--Paap $F$ statistic is 2{,}255.9 in every column. $^{*}\,p<0.1$, $^{**}\,p<0.05$, $^{***}\,p<0.01$.\par}
\end{table}

\end{document}